# Charge Carrier Transport in Iron Pyrite Thin Films: Disorder Induced Variable Range Hopping

*Sudhanshu Shukla*[1, 2], *Sinu Mathew*[3], *Hwan Sung Choe*[4, 5], *Manjusha Chugh*[6], *Thomas D. Kühne*[6], *Hossein Mirhosseini*[6], *Xiong Qihua*[7], *Junqiao Wu*[4,5,8], *Thirumalai Venkatesan*[3], *Thirumany Sritharan*[2], *Joel W. Ager*[4,5,8]

[1]Energy Research Institute and [2]School of Materials Science and Engineering, Nanyang Technological University, Singapore 637371, Singapore

[3]NUSNNI-NanoCore, National University of Singapore, Singapore 117576, Singapore

[4]Department of Materials Science and Engineering, University of California, Berkeley, California 94720, USA

[5]Materials Sciences Division, Lawrence Berkeley National Laboratory, Berkeley, California 94720, USA

[6]Dynamics of Condensed Matter and Center for Sustainable Systems Design, Chair of Theoretical Chemistry, University of Paderborn, Warburger Str. 100, D–33098 Paderborn, Germany

[7]Division of Physics and Applied Physics, School of Physical and Mathematical Sciences, Nanyang Technological University, Singapore 637371, Singapore

[8]Berkeley Educational Alliance for Research in Singapore (BEARS), Ltd., 1 Create Way, 138602, Singapore

**Abstract:** The origin of p-type conductivity and the mechanism responsible for low carrier mobility was investigated in pyrite (FeS$_2$) thin films. Temperature dependent resistivity measurements (10 – 400 K) were performed on polycrystalline and nanostructured thin films prepared by three different methods: (1) spray pyrolysis, (2) hot-injection synthesized and spin-coated nanocubes and (3) pulsed laser deposition. Films have a high hole density ($10^{18} – 10^{19}$) cm$^{-3}$ and low mobility (0.1 – 4 cm$^2$ V$^{-1}$ s$^{-1}$) regardless of the method used for their preparation. The charge transport mechanism is determined to be nearest neighbour hopping (NNH) at near room temperature with Mott-type variable range hopping (VRH) of holes *via* localized states occurring at lower temperatures. Density functional theory (DFT) predicts that sulfur vacancy induced localized defect states will be situated within the band gap with the charge remaining localized around the defect. The data indicate that the electronic properties including hopping transport in pyrite thin films can be correlated to sulfur vacancy related defect. The results provide insights on electronic properties of pyrite thin films and its implications for charge transport.

## I. Introduction:

Iron pyrite (FeS$_2$), often referred to as "fool's gold", has been of interest to the solar research community due to its suitable energy bandgap (~0.95 eV), large optical absorption coefficient (>10$^5$ cm$^{-1}$ for E > 1.3 eV), and the elemental abundance of Fe and S in the earth's crust.[1, 2] Bulk n-type single crystals can have high electron mobilities up to 360 cm$^2$ V$^{-1}$ s$^{-1}$ and long minority diffusion lengths in the range of 100 – 1000 nm.[1] However, attempts to make solar cells based on pyrite thin films have not been successful; the maximum reported solar to electrical conversion efficiency is <3 % with a AM1.5 photovoltage of only 200 mV.[3-6] There is a significant interest in understanding the charge transport mechanism in iron pyrite as it directly correlate to the fundamental issues responsible for poor performance such as – (i) presence of non-stoichiometric Fe-S conducting impurity phases (e.g., marcasite FeS$_2$, FeS, Fe$_3$S$_4$, Fe$_{1-x}$S), (ii) surface issues – reduced surface bandgap, surface inversion layer, non-stoichiometry due to reduced co-ordination, and Fermi level pinning at the surface. [4, 7-16] and (iii) intrinsic bulk defects. The last issue has been invoked to explain many commonly observed traits of pyrite thin films such as the lack of photoresponse, p-type conduction along with low hole mobility, and high dark currents.[14, 17] Therefore, study of the charge transport and doping behaviour of pyrite thin films is highly relevant in the context of any device applications of this material. Concerning charge transport, knowledge of temperature dependent resistivity is extremely valuable as it contains wealth of information regarding the conduction mechanism and electronic properties. Combining the charge transport information with charge carrier density and mobility values allows to understand the impact of defects and scattering processes within the material.

As evaluation of hopping-like conductivity is a main theme in this paper, we provide a brief overview of the basic transport theory. In disordered semiconductors, electronic conduction occurs via hopping through localized states. [18, 19] For charge conduction in disordered systems, the resistivity $\rho$ will vary characteristically with temperature $T$ as $\rho \propto \exp(T^{-\alpha})$, where α is a non-linearity exponent which depends on dimensionality of the system:

$$\rho = \rho_o * exp\ (T_o/T)^\alpha \qquad (1)$$

with $T_o$ being the characteristic temperature. The value of α = 1 correspond to the Arrhenius type thermally activated nearest-neighbour hopping conduction (NNH). The exponent α further distinguishes the variable range hopping (VRH) mechanism with α = 1/4 corresponding to Mott-VRH and α = 1/2 to Efros-Shklovskii-(ES) VRH. α = 1/2 is also related to granular conductivity through metallic clusters.[20] For the case of Mott- VRH the characteristic temperature is denoted as $T_M$ [21, 22]:

$$T_M = \frac{21.2}{k_B g_o \xi^3} \qquad (2)$$

where $g_o$ is the density of states (DOS) near the Fermi surface, $\xi$ is the carrier localization length and $k_B$ is the Boltzmann constant. The value of $g_o$ is used here for pyrite will be $8.5\times10^{19}$ cm$^{-3}$ eV$^{-1}$ [23]. Similarly, for ES-VRH conduction, the characteristic temperature $T_{ES}$ is related to the localization length $\xi$ (decay length of the exponentially localized wavefunctions) by the following relation [21, 22]

$$T_{ES} = \frac{2.8\ e^2}{4\pi\epsilon_o \kappa k_B \xi} \qquad (3)$$

where $\kappa$ is the dielectric constant (20.5),[24] and $e$ is the charge of an electron.

At high temperatures, we expect to have NNH type conduction behaviour. In this mechanism, carriers hop to their nearest neighbour site through thermally activated impurity conduction process. The functional dependence of ρ on T is given as $\rho = \rho_o \exp(E_A/k_B T)$, where $\rho_o$ is the prefactor ($\rho = \rho_o$ as T → ∞). Notably, α = 1 also describes the thermally activated band conduction which differs from NNH conduction, with the activation energy being much lower for the latter.

VRH conduction occurs *via* transport of carriers by hopping among the localized electronic states near the Fermi level and thus manifests the presence of structural disorder (random potential fluctuations in the periodic potential) in the material. In pyrite, the nanoscale stoichiometric deviation induced phase inhomogeneity and disorder are believed to cause exponentially decaying band tails and associated localized defect states/band.[9, 15] This is supported by the reported Mott-VRH conduction ($\alpha = ¼$) in n-type pyrite single crystals.[11, 25, 26] Efros-Shklovskii type hopping ($\alpha = ½$) is also observed in pyrite single crystals, where the Coulomb interaction leads to opening of an energy gap near the Fermi level.[11, 27] Regarding the nature of the defects responsible for hopping conduction, recent studies by Leighton and co-workers on n-type pyrite single crystals have established a direct correlation between sulfur vacancies and electronic charge transport mechanism, and further linked sulfur vacancy clusters to the experimentally observed defect activation energies in bulk crystals.[28, 29]

Prior work on determining the conduction mechanism in thin films has pointed to p-type thermally activated transport with low activation energies indicating the case of (NNH).[14, 30, 31] Mott-VRH transport has been reported in low mobility pyrite thin films prepared by *ex-situ* sulfurization of Fe by Zhang *et al*.[13, 31] In these films, the crossover from hopping conduction to thermally activated transport occurs at relatively higher temperatures than in the single crystals, indicating a more disordered system. In the similar study, Zhang *et al*. has explained an observation of $\alpha = ½$ in terms of inter-granular hopping (IGH) transport ascribed to un-reacted metallic Fe clusters in pyrite thin films akin to tunnelling transport mechanism.[31] Notably, $FeS_2$ formation from thermal sulfurization of Fe is a rate-limiting process which involves intermediate sulfur-deficient phases.[32, 33] Therefore the transport described above by Zhang *et al*. could be the characteristic of specific sample growth conditions rather than being an inherent property of pyrite thin films in general. Apart from that, a number of groups have endeavoured to control the synthesis conditions and prepare nanostructures of pyrite in order to gain insight into how the intrinsic carrier transport is affected by the presence of defects and phases impurities.[34-41] Mott-VRH conduction is also observed for nanostructured pyrite.[25] Polycrystalline thin film and nanostructures of pyrite show differences in the doping behaviour in comparison to single crystals. Thin films and nanostructured pyrite are mostly reported to be p-type with low carrier mobility while the single crystals are n-type with high carrier mobility. Carrier mobility values in single crystals are argued to be limited by ionized impurity scattering.[28] Considering a disordered system, low mobility polycrystalline thin films and nanostructures need to be scrutinized for both intrinsic electron-phonon scattering and ionized impurity scattering mechanism. While much has been discussed regarding the nature and origin of electronic defects in single crystals, there is still no consensus on the interpretation of carrier transport and doping mechanism in polycrystalline and nanocrystalline p-type thin films and the impact of defect and phase impurity.[14, 30, 42]

Since the formation of intrinsic defects depends on the growth conditions, analysis of pyrite thin films by combining various synthesis strategies would provide in-depth understanding of these issues. To this end, we prepared pyrite thin films using three different methods in order to identify and examine the influence of intrinsic factors on charge transport and shed light on the nature of associated defects. (1) Spray pyrolysis, while inexpensive and scalable, offers less control of stoichiometry and crystal quality compared to nanocrystal synthesis and vacuum based deposition processes. On the other hand, (2) hot-injection synthesis might provide more control of the shape, size and stoichiometry and hence better film quality. (3) Vacuum-based deposition *via* pulsed laser deposition (PLD), provides excellent control of the film thickness and can avoid impurities if a suitably pure target is used. Therefore, here we used a natural pyrite crystal of high electronic quality directly as the target for PLD film deposition. Details of the synthesis procedure and sample preparation are described in supplementary information.

We note that the conductivity type determination in pyrite thin films require care in light of the low hole mobility and variable range hopping transport. Therefore, our experimental design includes both Hall and thermoelectric measurements to substantiate our conclusions regarding the determination of charge carrier type.

## II. Results and Analysis:

### A. Phase and morphology analysis

Prior to evaluation of the transport mechanism, we analysed the morphology and phase purity for all the films in the study. Figure 1 (a) compares film morphology. Polycrystalline spray pyrolyzed thin films consist of columnar grains, while PLD films are smooth and compact with small grain sizes. Thin films made from hot-injection synthesis and sulfurization consist of compact randomly oriented nanocubes. Figure 1 (b) shows the Raman spectra of the pyrite films synthesized by different methods. All films show peaks around 342 cm$^{-1}$ and 378 cm$^{-1}$, which correspond to $A_g$ and $E_g$ vibrational modes characteristic of cubic pyrite phase. As-grown PLD films were amorphous; therefore PLD deposited films were sulfurized in (1) sulfur vapour and by (2) an $H_2S$ plasma to obtain crystalline and phase pure $FeS_2$. $H_2S$ plasma sulfurization of iron oxide thin film has been reported to be more effective in achieving $FeS_2$ compared to thermal sulfurization from sulfur powder by avoiding sulfur deficient phases during the sulfurization process.[43] On the other hand, thermal sulfurization proceeds through sulfur deficient phases that could limit the sulfur diffusion thereby inhibiting complete elimination of sulfur deficient phases. However, an additional peak at 326 cm$^{-1}$ corresponding to marcasite phase is observed for the $H_2S$ plasma sulfurized PLD deposited film (PLD film-2).

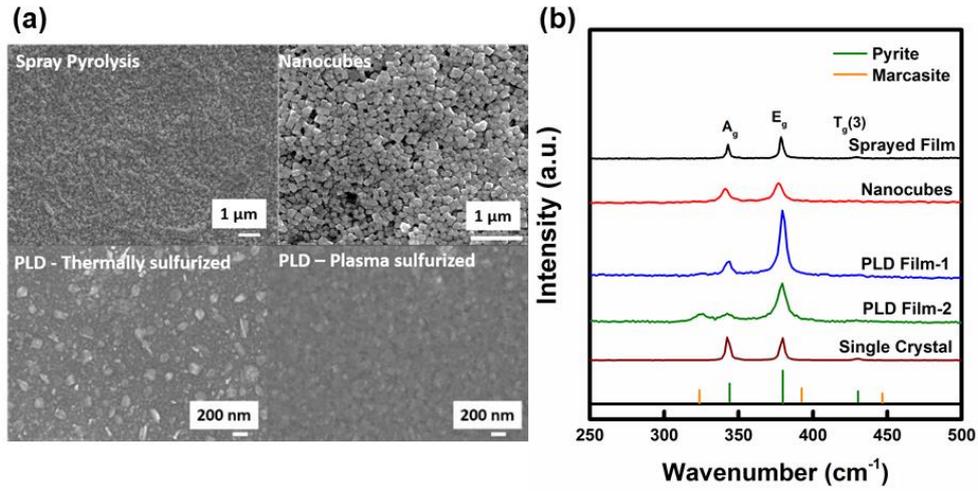

**Figure 1.** (a) top-view SEM micrographs showing the morphology of the pyrite thin films prepared by spray pyrolysis, hot-injection formation of nanocubes followed by spin coating, and pulsed laser deposition. PLD film-1 and PLD film-2 refer to the thermally and $H_2S$ plasma sulfurized PLD films respectively. (b) Raman spectra of the films shown in (a).

**B. Temperature dependent electrical transport measurements**

For temperature dependent transport measurements aluminium metal contacts (work function close to $E_F$ of pyrite) were deposited at the corners of the film and the measurements were performed in van der Pauw (lateral) configuration. Figure 2 shows the temperature dependent electronic transport data for all pyrite thin films analysed in this report. All the films have increasing resistivity with decreasing temperature, as expected for activated transport in a disordered semiconductor.

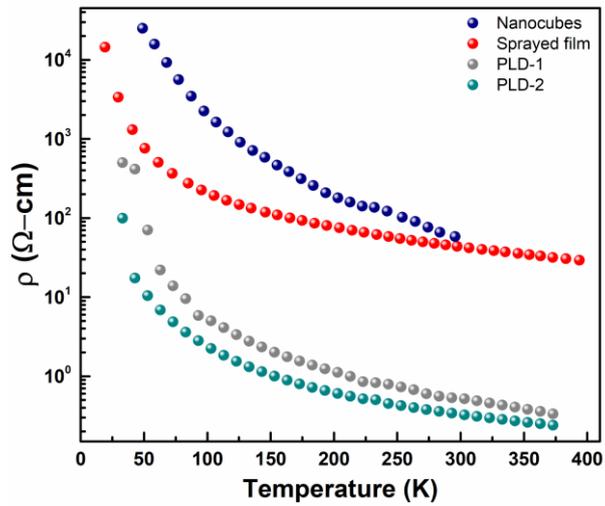

**Figure 2.** Temperature dependence of resistivity ($\rho$ vs $T$) for pyrite thin films prepared by spray pyrolysis, hot-injection formation of nanocubes followed by spin coating, and pulsed laser deposition.

Regimes of particular conduction mechanism are differentiated based on their characteristic dependence of resistivity on temperature as discussed in the Introduction. For all experimental data, the determination of the conductivity regime was further augmented by linearization analysis from reduced activation energy plots (Zabrodskii plots), as discussed in the supplementary information (section 6, Figure S3). [44]

**Spray pyrolyzed films.** For sprayed pyrite thin films, temperature dependent resistivity was measured in the range of 400 K – 10 K. Evidence for all three transport mechanisms: Evidence for Arrhenius/NNH, Mott-VHR, and Efros-Shklovskii-(ES) VRH was found, as shown in Figure 3.

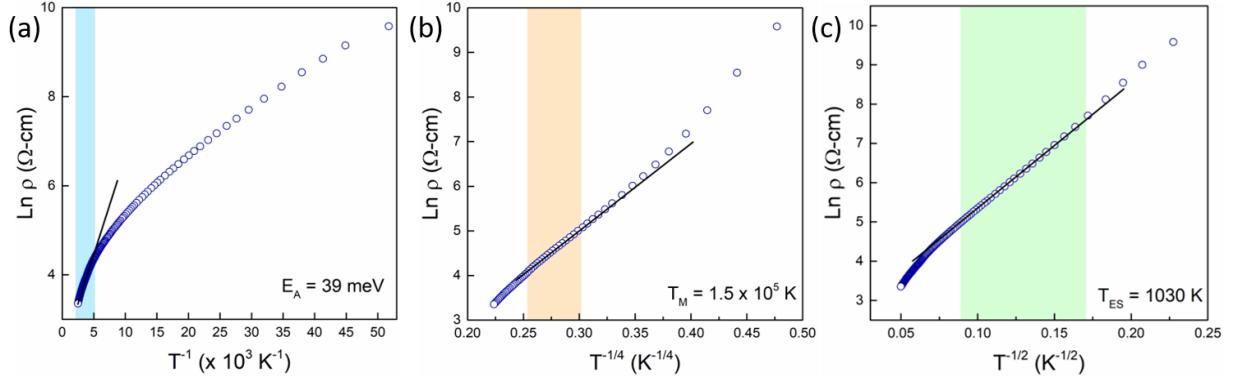

**Figure 3.** (a) Ln $\rho$ *vs* $1/T$, and (b) Ln $\rho$ *vs* $1/T^{1/4}$ and (c) Ln $\rho$ *vs* $1/T^{1/2}$ plots for spray pyrolyzed pyrite thin film. Shaded regions in (a), (b) and (c) depicts NNH, VRH and ES-VRH conduction range respectively.

The shaded area in Figure 3(a) indicates the regime (230 – 400 K) of Arrhenius type thermally activated transport (conventional transport through thermal diffusion, TAC) with a small activation energy $E_A$ of 39 meV. At lower temperatures, Figure 3(b), the experimental data are fit well by Eq. 1 with the exponent value α = ¼, indicating a Mott-VRH conduction mechanism in the wide temperature range of 120 – 220 K. From the slope obtained after the fitting with Mott-VRH model, the characteristic temperature $T_M$ is determined to be 1.52 x 10$^5$ K. Using Eq. (2) and the value of $T_M$, localization length (ξ) is calculated to be around 2.68 nm. Upon further lowering the temperature, conduction follows α = ½ relation in the range ~ 32 – 125 K corresponding to the ES-VRH mechanism. The localization length (ξ) calculated from Eq. (3) is 2.1 nm, close to the value obtained from Mott-VRH hopping, which suggests that similar defects participate in the conduction for the two mechanisms.

Variable range hopping observed over wide temperature range indicates the high degree of disorder in pyrite thin films made by spray pyrolysis. As the temperature is increased from the ES to Mott regime, the Coulomb gap vanishes and a crossover to Mott-type conduction is observed. The theoretical crossover temperature calculated by using the expression below[21]

$$T_{crossover} = 16 \left( \frac{T_{ES}^2}{T_{Mott}} \right) \qquad (4)$$

gives T$_{crossover}$ ~ 111 K, consistent with the experimentally observed crossover temperature.

**Nanocubes**. For the nanocubes film, the transport data is shown in Figure 4. Thermally activated transport is observed from ~ 240 – 300 K with activation energy of 87 meV (Figure 4 (a)). This value is two times higher than the value obtained for spray-pyrolyzed films which have overall lower resistivity but is close to a value obtained previously for pyrite nanostructures.[25] This indicates that the defect states lie deeper in the gap in the case of nanocubes as compared to the sprayed films. We interpret the difference in activation energies in terms of higher defect density in sprayed films. For higher defect densities, the defect levels are broaden forming a defect band and shift the Fermi level position and hence a reduced activation energy could be obtained. Mott-VRH type conduction is observed over a wide temperature range 230 – 65 K with a high characteristic temperature $T_M$ of ~ 4.5 x 10$^6$ K (Figure 4 (b)). Further decreasing the temperature resulted in resistivity values which were too high to be accurately analysed.

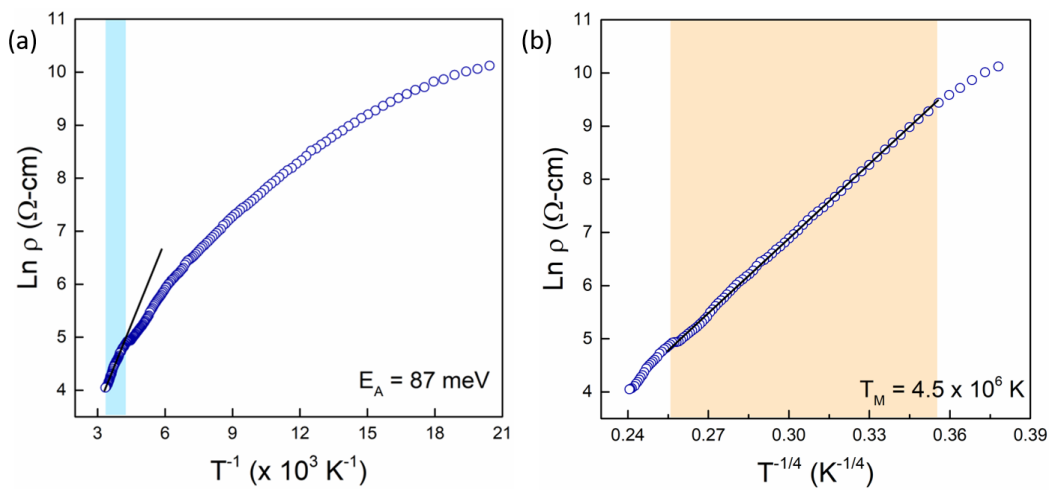

**Figure 4.** (a) Ln $\rho$ vs 1/$T$ and (b) Ln $\rho$ vs 1/$T^{1/4}$ plot for the spin-coated film made from nanocubes. Shaded regions in (a) and (b) depicts thermally activated and VRH conduction regime respectively.

This high value of $T_M$ is indicative of a highly disordered system. Mott-VRH conduction was previously observed by Acevedo *et al.* in pyrite nanostructures with similar high value of $T_M$.[25] The characteristic length scales are claimed to be small (Bohr radius ~ 2.3 nm) in pyrite to observe the effect of quantum confinement and dimensionality.[25, 45] The carrier localization length from Eq. (2) is calculated to be ~ 0.86 nm. Higher activation energy and lower value of localization length reinforces the argument of deeper defect state.

**PLD films.** Figure 5 (a) shows that activated NNH type transport occurs in the films with close values of activation energies in the temperature range 260 – 370 K. Zabrodskii plots (Figure S3) indicate a mixed conduction regime of thermally activated and hopping transport from 260 – 290 K. As shown in Figure 5 (b) and similar to the other two classes of thin films, thermal and plasma sulfurized film show Mott-VRH conduction in the temperature range 40 – 290 K with high Mott characteristic temperatures $T_M$ of 8.1 x $10^5$ K and 5.3 x $10^5$ K respectively. Localization lengths for thermally and plasma sulfurized film are determined to be ~1.52 nm and 1.76 nm respectively. The mixed phase film (PLD film-2) was slightly more conducting than the phase pure film.

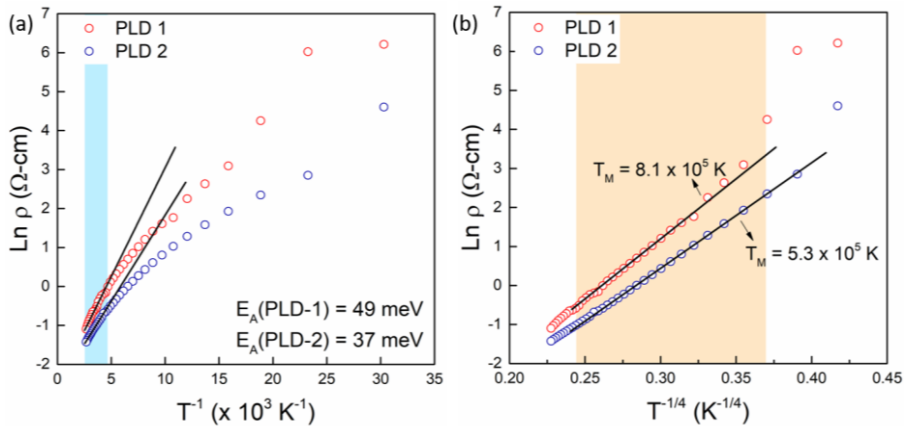

**Figure 5.** (a) Ln $\rho$ vs $1/T$ and (b) Ln $\rho$ vs $1/T^{1/4}$ plots for PLD deposited thermally sulfurized and H$_2$S plasma sulfurized film. Shaded regions in (a) and (b) depicts thermally activated and VRH conduction regime respectively.

Parameters obtained from the analysis of the resistivity data are summarized in Table I. Also included in Table I are room temperature Hall effect and Seebeck coefficient data.

**Table 1**. Electrical characteristics of $FeS_2$ studied in this report.

| Parameters | Spray pyrolyzed film | Nanocubes | Thermally sulfurized PLD film | Plasma sulfurized PLD film | Natural Crystal |
|---|---|---|---|---|---|
| Resistivity (Ω-cm) | 0.54 | 11.20 | 0.54 | 0.37 | 1.8 |
| Mobility (cm²/V-s) | 4.22 | 0.10 | 1.4 | 0.3 | 64.9 |
| Carrier density (cm⁻³) | $2.7 \times 10^{18}$ | $5.3 \times 10^{18}$ | $7 \times 10^{18}$ | $4 \times 10^{19}$ | $5.1 \times 10^{16}$ |
| Carrier type (Hall/Seebeck) | p/+35 µV/K | p/- | p/ +74 µV/K | p/ +64 µV/K | n/- |
| Bandgap (eV) | 1.04 | 1.02 | 1.01 | 1.05 | 1.2 |
| Conduction mechanism | TAC (> 230 K) Mott-VRH (120-220 K) ES-VRH (32-125 K) | TAC (> 240 K) VRH (65-230 K) | TAC (> 260 K) VRH (60-290 K) | TAC (> 260 K) VRH (40-290 K) | - |
| Activation energy ($E_A$) | 39 meV | 87 meV | 49 meV | 37 meV | |
| Localization length ($\xi$) | 2.7 nm | 0.9 nm | 1.5 nm | 1.8 nm | |

The carrier concentration in pyrite thin films obtained after different synthesis approach does not vary significantly with preparation method and is in the range $\sim 10^{18}$ cm⁻³ with low carrier mobility. As expected, the films had positive Hall and Seebeck coefficients indicating p-type conductivity while the natural crystal was n-type. All pyrite films studied in this paper, regardless of preparation method show similar VRH conduction and small localization lengths indicating high density of localized defect states (Table 1). Thus, the impact of intrinsic disorder in pyrite is the significant factor that severely affects the electrical properties of pyrite thin films. Localization length in the range of 0.8 – 2 nm for all the samples reflects the short-range order.

**C. Theoretical calculations –**

Sulfur vacancies are predicted to induce mid-gap states where the carriers can be localized [10, 45-47], motivating us to perform DFT calculations to determine their energy position in the gap. For pristine FeS$_2$, calculated bulk lattice parameter of iron pyrite (FeS$_2$) is found to be 5.428 Å, which is close to the experimental value (5.416 Å). The indirect bulk band gap within PBE+U was calculated to be 0.99 eV, which is close to the experimental value (0.95 eV) (Supplementary Figure S4). Fig.6 (a) and (b) shows spin-resolved density of states (DOS) of bulk defect-free and sulfur vacancy defect containing FeS$_2$. The DOS plot clearly shows that non-ferromagnetic nature of bulk FeS$_2$. In both case, the top of the valence band is dominated by Fe-3d orbitals while bottom of the conduction band is formed by hybridization of S-3p and Fe-3d orbitals, S-3p being dominated at Gamma (Γ) point. In order to model a sulfur vacancy (SV), a 2×2×2 unit cell containing 96 atoms was considered. The spin-resolved DOS for a unit cell containing a single SV is shown in Fig. 6(b). A single SV generates three states in the band gap, consistent with the previous theoretical studies.[29, 45]. Two of these are close to valence band maximum (VBM) and one is very close to conduction band minimum (CBM) which is a resonant state with CBM. These states originates because of broken bonds of Fe and S atom near to the SV. The states near to VBM are dispersionless and occupied.

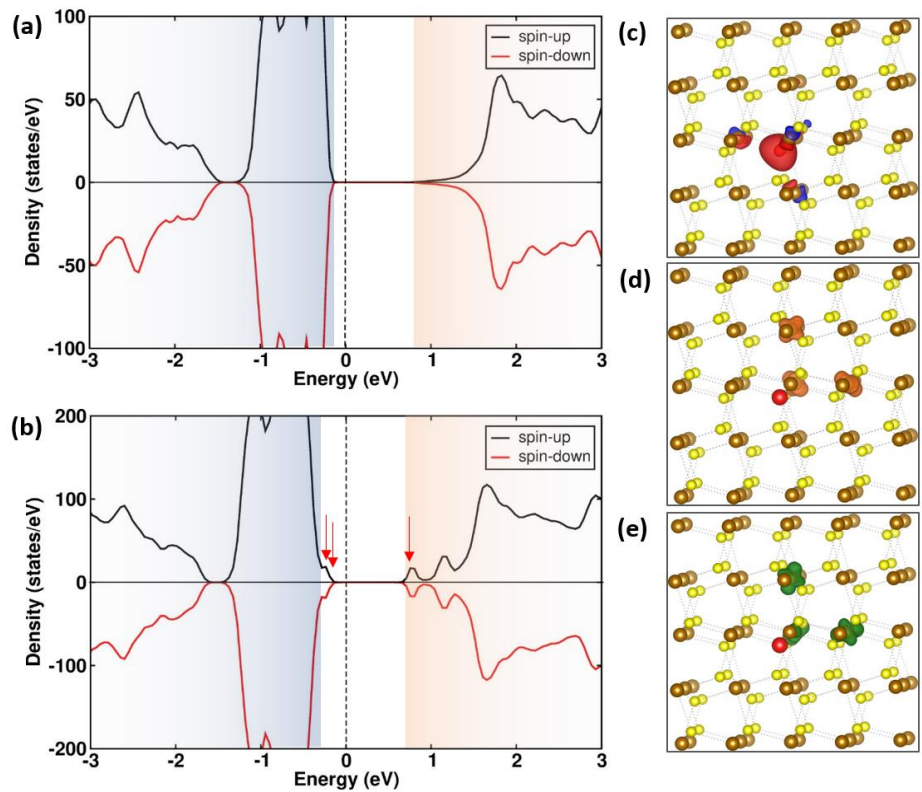

**Figure 6.** Calculated total DOS of (a) bulk $FeS_2$ and (b) $FeS_2$ containing a sulfur vacancy (SV). Both spin-up and spin-down DOS are shown. Red arrows indicate the defect states. (c) Isosurface ($\rho$=0.085e/Bohr$^3$) of total charge density difference around the SV. Red color indicates charge accumulation and blue color indicates charge depletion. Isosurface ($\rho$=0.015e/Bohr$^3$) of decomposed charge density for (d) first defect and (e) second defect state close to VBM in a 2×2×2 unit cell containing a SV. The location of SV is shown as a red coloured sphere. Brown and yellow coloured spheres represent Fe and S atoms, respectively.

In order to estimate the extent of charge distribution around SV, we calculated the charge distribution around the SV. Fig. 6 (c) shows the isosurface of charge density difference arising because of SV. Most of the excess charge is accumulated around the SV defect. From these charge density distribution plots, it can be seen that the charge perturbation around a single vacancy point defect remains localized within few angstroms range ($\approx$ 3-6 Å) around the defect. The experimentally observed localization length from transport measurements in the range of ~ 1 – 2 nm could be reconciled with vacancy clustering, which is shown highly likely to exist in $FeS_2$ single crystals.[29, 48] We expect similar to happen in thin films as well. Our previous magnetic measurements on pyrite nanocubes also provided indirect evidence of clustering of defects in pyrite.[9] The electronic behavior of the defect state is difficult to predict, for example – an occupied neutral sulfur vacancy defect near VBM is likely to act as donor while positively charged defect is likely to act as acceptor. Bader charge analysis[49] of defect-free and a system containing a SV shows that – while partial charge on Fe atoms remain same (+0.91e), those on S atoms changes. In defect-free unit cell, S atoms of the S-S dimer have partial charges of -0.42e and -0.48e. In case of a single SV, S atoms away from the vacancy site still have the same partial charges but the S atom of the broken S-S dimer bond has a partial charge of -0.76e. This implies that the charge remains localized near to the SV. Similar decomposed charge density distributions for valence band maximum (VBM) and conduction band minimum (CBM) is shown in supplementary information (Fig. S4).

**III. Discussion**

Variable range hopping conduction has been previously observed and was seen also in this study for low mobility and p-type pyrite thin films and nanostructures. We also observed, as have prior studies, that single crystals can have relatively high carrier mobility and are n-type. In Figure 7 (a), we plot the carrier concentration *versus* mobility values for pyrite samples reported in the literature along with our data of pyrite thin films studied from the Hall measurements to understand the carrier scattering mechanism. Notable observations from the plot are: (i) Carrier concentration varies over a wide range even for unintentionally doped pyrite samples, (ii) Mobility decreases over four orders of magnitude as carrier density increases from $10^{16}$ cm$^{-3}$ – $10^{21}$ cm$^{-3}$, and most strikingly (iii) high mobility samples, mostly single crystals, show n-type conduction while thin film and nanostructures exhibit p-type behaviour with low mobility. (Regimes classified in two different shaded region). Thin film samples studied in this work lie in the high carrier concentration and low mobility p-type regime, suggestive of higher defect concentration in thin films compared to single crystals.

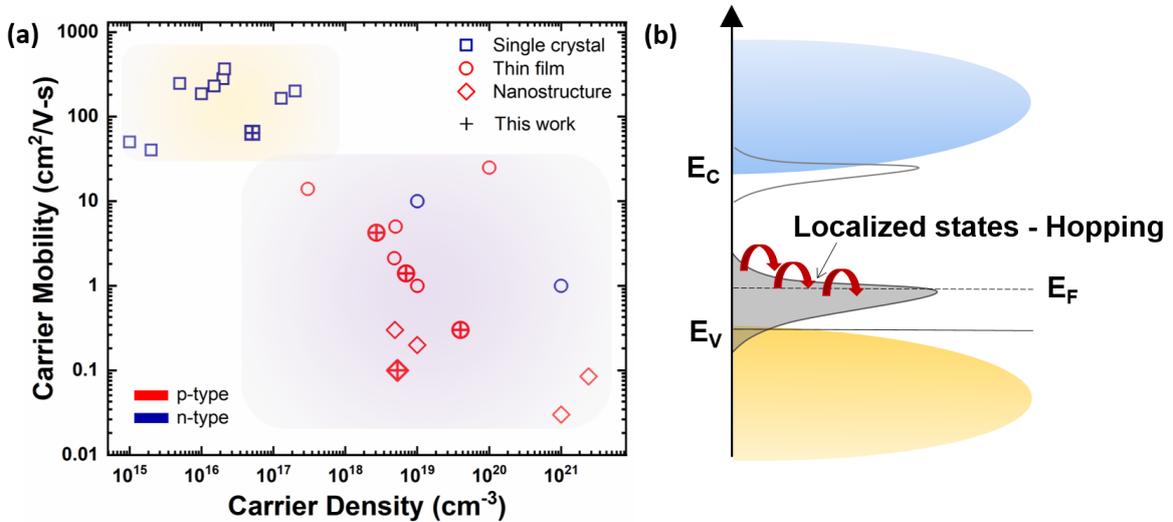

**Figure 7.** (a) Room temperature Hall mobility *vs* carrier concentration graph for pyrite single crystal and thin films reported in literature and samples studied in the current work. Data points in the plot are taken from the references.[4, 11, 12, 25, 26, 28, 30, 31, 36, 50-57] (b) DOS and defect band close to valence band and conduction band is shown. Fermi level $E_F$ lies in the vicinity of defect band where carrier hopping is depicted.

The lower carrier mobility with high carrier density is a consequence of the progressive spatial localization of the electronic density. Higher concentration of intrinsic impurity defects result in enhanced carrier scattering and thus limit the hole mobility in p-type films. A contrary argument against impurity scattering limited mobility could be given in terms of enhanced screening effects with increased carrier densities in the presence of compensated background charged defects. Notably, Leighton *et al.* observed higher maximum carrier mobility ($\mu_{max}$) with high electron density for sulfur deficient pyrite crystal than sulfur-rich sample with lower electron density. They attributed this to enhanced screening in high carrier density sample.[28] Thus, high carrier concentration and low mobility observed in thin films cannot be reconciled with the screening phenomenon but can be explained by an unavoidable and intrinsic mobility limiting process stemming from lattice scattering mechanism *i.e.* enhanced phonon scattering.[58] We explain lower mobilities in our films in terms of phonon scattering. In the presence of localized states, spatial localization of carriers allows uncertainty in the momentum values and involves higher phonon contribution.[59] This is directly evidenced by our observation of carrier localization and phonon assisted VRH conduction with small localization lengths and low activation energies derived from NNH.

Based on the results obtained from experimental and theoretical analysis, we attempt to draw a physical picture as shown in Figure 7 (b). High density of defects lead to the formation of localized states within the bandgap. VRH conduction occurs through carrier hopping across these localized states close to the Fermi level ($E_F$). Sulfur vacancy defects, particularly sulfur vacancy clusters,[29] which are most often invoked in the literature to be responsible for defect states within the bandgap, might explain the observed transport behaviour of pyrite thin films similar to the single crystals. Localization lengths in few nanometer scale provides evidence in support of sulfur vacancies that are directly linked to the localized states. High hole concentration, low carrier mobility and VRH conduction are consistent with this picture. We find the variable range hopping transport in the pyrite thin films regardless of preparation method, indicating a highly disordered semiconductor with high intrinsic defect density. Hopping conduction similar to the films studied by Zhang *et al*. are obtained for all the studied films regardless of the synthesis method adopted for the growth, implying that intrinsic factors predominantly govern properties.

**IV. Conclusion:**

In summary, we have investigated the electrical charge transport in various pyrite films. All the films exhibit Mott-VRH conduction over a wide temperature range irrespective of the film preparation method. In addition to that, films showed p-type conductivity with high hole concentration along with low mobility and low thermal activation energy of around 39, 87 and 49 and 37 meV for sprayed, nanocubes, thermally sulfurized PLD (PLD-1) and plasma sulfurized (PLD-2) deposited film respectively, exhibiting nearest neighbour hopping transport at room temperature. Mott to ES- VRH crossover is observed in spray pyrolyzed thin film. The hopping transport model is described by the conduction though localized states in the impurity band and the corresponding calculated localization length could be correlated to sulfur vacancies consistent with the DFT calculations of band structure after introducing sulfur vacancy defect. We also show that decreased mobility of pyrite thin films compared to single crystals can be explained by phonon scattering mechanism which is expected for the case of strong localization of carriers. Thus, the electronic behaviour of pyrite is strongly affected by the disorder induced by intrinsic defects.

Hence, strategies to pursue different methods for depositing pyrite thin films will not improve the situation unless the localized defect states are eliminated by mitigating the spontaneous formation of sulfur vacancies. This requires taming of intrinsic sulfur vacancy defects and controlling the degree of disorder gain better control over electronic properties of pyrite. Potential approach that could be useful in this direction is to push the reaction kinetics in the sulfur rich domain throughout the growth process or develop passivation strategy for such defects. This would help in manipulating electronic properties of pyrite and subsequently enabling improved device performance.

**Acknowledgments**

This work was supported by the National Research Foundation (NRF) through the Singapore-Berkeley Research Initiative for Sustainable Energy (SinBeRISE) CREATE Programme. Seebeck coefficient measurements, Hall measurements and PLD of pyrite thin films were performed in the Electronic Materials Program, which is supported by the Director, Office of Science, Office of Basic Energy Sciences, Materials Sciences and Engineering Division of the U.S. Department of Energy under contract no. DE-AC02-05CH11231. M. Chugh would like to acknowledge the Paderborn Center for Parallel Computing (PC2) for computing time on OCuLUS and FPGA-based supercomputer NOCTUA.

**References**

1. Ennaoui, A., et al., *Iron disulfide for solar energy conversion.* Solar Energy Materials and Solar Cells, 1993. **29**(4): p. 289-370.
2. Wadia, C., A.P. Alivisatos, and D.M. Kammen, *Materials Availability Expands the Opportunity for Large-Scale Photovoltaics Deployment.* Environmental Science & Technology, 2009. **43**(6): p. 2072-2077.
3. Ennaoui, A. and H. Tributsch, *Iron sulphide solar cells.* Solar Cells, 1984. **13**(2): p. 197-200.
4. Cabán-Acevedo, M., et al., *Ionization of High-Density Deep Donor Defect States Explains the Low Photovoltage of Iron Pyrite Single Crystals.* Journal of the American Chemical Society, 2014. **136**(49): p. 17163-17179.
5. Rahman, M.Z. and T. Edvinsson, *What Is Limiting Pyrite Solar Cell Performance?* Joule, 2019.

6. Steinhagen, C., et al., *Pyrite Nanocrystal Solar Cells: Promising, or Fool's Gold?* The Journal of Physical Chemistry Letters, 2012. **3**(17): p. 2352-2356.
7. Ennaoui, A. and H. Tributsch, *Energetic characterization of the photoactive FeS2 (pyrite) interface.* Solar Energy Materials, 1986. **14**(6): p. 461-474.
8. Herbert, F.W., et al., *Quantification of electronic band gap and surface states on FeS2(100).* Surface Science, 2013. **618**: p. 53-61.
9. Shukla, S., et al., *Origin of Photocarrier Losses in Iron Pyrite (FeS2) Nanocubes.* ACS Nano, 2016. **10**(4): p. 4431-4440.
10. Bronold, M., Y. Tomm, and W. Jaegermann, *Surface states on cubic d-band semiconductor pyrite (FeS2).* Surface Science, 1994. **314**(3): p. L931-L936.
11. Limpinsel, M., et al., *An inversion layer at the surface of n-type iron pyrite.* Energy & Environmental Science, 2014. **7**(6): p. 1974-1989.
12. Liang, D., et al., *Gated Hall Effect of Nanoplate Devices Reveals Surface-State-Induced Surface Inversion in Iron Pyrite Semiconductor.* Nano Letters, 2014. **14**(12): p. 6754-6760.
13. Zhang, X., et al., *Phase Stability and Stoichiometry in Thin Film Iron Pyrite: Impact on Electronic Transport Properties.* ACS Applied Materials & Interfaces, 2015. **7**(25): p. 14130-14139.
14. Seefeld, S., et al., *Iron Pyrite Thin Films Synthesized from an Fe(acac)3 Ink.* Journal of the American Chemical Society, 2013. **135**(11): p. 4412-4424.
15. Lazić, P., et al., *Low intensity conduction states in FeS 2 : implications for absorption, open-circuit voltage and surface recombination.* Journal of Physics: Condensed Matter, 2013. **25**(46): p. 465801.
16. Shukla, S., et al., *Scientific and Technological Assessment of Iron Pyrite for Use in Solar Devices.* Energy Technology, 2018. **6**(1): p. 8-20.
17. Shukla, S., et al., *Iron Pyrite Thin Film Counter Electrodes for Dye-Sensitized Solar Cells: High Efficiency for Iodine and Cobalt Redox Electrolyte Cells.* ACS Nano, 2014. **8**(10): p. 10597-10605.
18. Ambegaokar, V., B.I. Halperin, and J.S. Langer, *Hopping Conductivity in Disordered Systems.* Physical Review B, 1971. **4**(8): p. 2612-2620.
19. Shore, K.A., *Electronic Processes in Non-crystalline Materials (Second Edition), by N.F. Mott and E.A. Davis.* Contemporary Physics, 2014. **55**(4): p. 337-337.
20. Nandi, U., D. Jana, and D. Talukdar, *Scaling description of non-ohmic direct current conduction in disordered systems.* Progress in Materials Science, 2015. **71**: p. 1-92.
21. Liu, H., A. Pourret, and P. Guyot-Sionnest, *Mott and Efros-Shklovskii Variable Range Hopping in CdSe Quantum Dots Films.* ACS Nano, 2010. **4**(9): p. 5211-5216.
22. Gantmakher, V.F. and L.I. Man, *Electrons and disorder in solids.* 2005.
23. Altermatt, P.P., et al., *Specifying targets of future research in photovoltaic devices containing pyrite (FeS2) by numerical modelling.* Solar Energy Materials and Solar Cells, 2002. **71**(2): p. 181-195.
24. Choi, S.G., et al., *Pseudodielectric function and critical-point energies of iron pyrite.* Physical Review B, 2012. **86**(11): p. 115207.
25. Cabán-Acevedo, M., et al., *Synthesis, Characterization, and Variable Range Hopping Transport of Pyrite (FeS2) Nanorods, Nanobelts, and Nanoplates.* ACS Nano, 2013. **7**(2): p. 1731-1739.
26. Zhang, X., et al., *Potential resolution to the doping puzzle in iron pyrite: Carrier type determination by Hall effect and thermopower.* Physical Review Materials, 2017. **1**(1): p. 015402.
27. Walter, J., et al., *Surface conduction in n-type pyrite FeS$_2$ single crystals.* Physical Review Materials, 2017. **1**(6): p. 065403.

28. Voigt, B., et al., *Transport Evidence for Sulfur Vacancies as the Origin of Unintentional n-Type Doping in Pyrite FeS2.* ACS Applied Materials & Interfaces, 2019. **11**(17): p. 15552-15563.
29. Ray, D., et al., *Sulfur Vacancy Clustering and Its Impact on Electronic Properties in Pyrite FeS2.* Chemistry of Materials, 2020. **32**(11): p. 4820-4831.
30. Berry, N., et al., *Atmospheric-Pressure Chemical Vapor Deposition of Iron Pyrite Thin Films.* Advanced Energy Materials, 2012. **2**(9): p. 1124-1135.
31. Zhang, X., et al., *Crossover From Nanoscopic Intergranular Hopping to Conventional Charge Transport in Pyrite Thin Films.* ACS Nano, 2013. **7**(3): p. 2781-2789.
32. Pimenta, G. and W. Kautek, *Pyrite film formation by H2S reactive annealing of iron.* Thin Solid Films, 1994. **238**(2): p. 213-217.
33. Pimenta, G. and W. Kautek, *Thermodynamic aspects of pyrite film formation by sulphur conversion of iron.* Thin Solid Films, 1992. **219**(1): p. 37-45.
34. Macpherson, H.A. and C.R. Stoldt, *Iron Pyrite Nanocubes: Size and Shape Considerations for Photovoltaic Application.* ACS Nano, 2012. **6**(10): p. 8940-8949.
35. Lucas, J.M., et al., *Ligand-Controlled Colloidal Synthesis and Electronic Structure Characterization of Cubic Iron Pyrite (FeS2) Nanocrystals.* Chemistry of Materials, 2013. **25**(9): p. 1615-1620.
36. Cabán-Acevedo, M., et al., *Synthesis and Properties of Semiconducting Iron Pyrite (FeS2) Nanowires.* Nano Letters, 2012. **12**(4): p. 1977-1982.
37. Bi, Y., et al., *Air Stable, Photosensitive, Phase Pure Iron Pyrite Nanocrystal Thin Films for Photovoltaic Application.* Nano Letters, 2011. **11**(11): p. 4953-4957.
38. Bai, Y., et al., *Universal Synthesis of Single-Phase Pyrite FeS2 Nanoparticles, Nanowires, and Nanosheets.* The Journal of Physical Chemistry C, 2013. **117**(6): p. 2567-2573.
39. Gong, M., A. Kirkeminde, and S. Ren, *Symmetry-Defying Iron Pyrite (FeS2) Nanocrystals through Oriented Attachment.* Scientific Reports, 2013. **3**: p. 2092.
40. Puthussery, J., et al., *Colloidal Iron Pyrite (FeS2) Nanocrystal Inks for Thin-Film Photovoltaics.* Journal of the American Chemical Society, 2011. **133**(4): p. 716-719.
41. Ge, H., et al., *Evolution of nanoplate morphology, structure and chemistry during synthesis of pyrite by a hot injection method.* RSC Advances, 2014. **4**(32): p. 16489-16496.
42. Wu, L., et al., *Enhanced Photoresponse of FeS2 Films: The Role of Marcasite–Pyrite Phase Junctions.* Advanced Materials, 2016. **28**(43): p. 9602-9607.
43. Morrish, R., R. Silverstein, and C.A. Wolden, *Synthesis of Stoichiometric FeS2 through Plasma-Assisted Sulfurization of Fe2O3 Nanorods.* Journal of the American Chemical Society, 2012. **134**(43): p. 17854-17857.
44. Zabrodskii, A.G., *The Coulomb gap: The view of an experimenter.* Philosophical Magazine B, 2001. **81**(9): p. 1131-1151.
45. Hu, J., et al., *First-principles studies of the electronic properties of native and substitutional anionic defects in bulk iron pyrite.* Physical Review B, 2012. **85**(8): p. 085203.
46. Aravind, K., et al., *Electronic states of intrinsic surface and bulk vacancies in FeS 2.* Journal of Physics: Condensed Matter, 2013. **25**(4): p. 045004.
47. Birkholz, M., et al., *Sulfur deficiency in iron pyrite ($FeS_{2-x}$) and its consequences for band-structure models.* Physical Review B, 1991. **43**(14): p. 11926-11936.
48. Herbert, F.W., et al., *Dynamics of point defect formation, clustering and pit initiation on the pyrite surface.* Electrochimica Acta, 2014. **127**: p. 416-426.

49. Henkelman, G., A. Arnaldsson, and H. Jónsson, *A fast and robust algorithm for Bader decomposition of charge density.* Computational Materials Science, 2006. **36**(3): p. 354-360.
50. Schieck, R., et al., *Electrical properties of natural and synthetic pyrite (FeS2) crystals.* Journal of Materials Research, 1990. **5**(7): p. 1567-1572.
51. Willeke, G., et al., *Thin pyrite (FeS2) films prepared by magnetron sputtering.* Thin Solid Films, 1992. **213**(2): p. 271-276.
52. Willeke, G., et al., *Preparation and electrical transport properties of pyrite (FeS2) single crystals.* Journal of Alloys and Compounds, 1992. **178**(1): p. 181-191.
53. Mazón-Montijo, D.A., M.T.S. Nair, and P.K. Nair, *Iron Pyrite Thin Films via Thermal Treatment of Chemically Deposited Precursor Films.* ECS Journal of Solid State Science and Technology, 2013. **2**(11): p. P465-P470.
54. Takahashi, N., et al., *Growth of Single-Crystal Pyrite Films by Atmospheric Pressure Chemical Vapor Deposition.* Chemistry of Materials, 2003. **15**(9): p. 1763-1765.
55. Echarri, A.L. and C. Sánchez, *"n" type semiconductivity in natural single crystals of FeS2 (pyrite).* Solid State Communications, 1974. **15**(5): p. 827-831.
56. Lichtenberger, D., et al., *Structural, optical and electrical properties of polycrystalline iron pyrite layers deposited by reactive d.c. magnetron sputtering.* Thin Solid Films, 1994. **246**(1): p. 6-12.
57. Kinner, T., et al., *Majority Carrier Type Control of Cobalt Iron Sulfide (CoxFe1–xS2) Pyrite Nanocrystals.* The Journal of Physical Chemistry C, 2016. **120**(10): p. 5706-5713.
58. Rahman, M., et al., *On the Mechanistic Understanding of Photovoltage Loss in Iron Pyrite Solar Cells.* Advanced Materials, 2020. **32**(26): p. 1905653.
59. Farvacque, J.L. and Z. Bougrioua, *Carrier mobility versus carrier density in $Al_xGa_{1-x}N/GaN$ quantum wells.* Physical Review B, 2003. **68**(3): p. 035335.

# Supplementary Information

# Charge Carrier Transport in Iron Pyrite Thin Films: Disorder Induced Variable Range Hopping

*Sudhanshu Shukla, Sinu Mathews, Hwan Sung Choe, Manjusha Chugh, Thomas D. Kühne, Hossein Mirhosseini, Xiong Qihua, Wu Junqiao, T. Venkatesan, W. Walukiewicz, Thirumany Sritharan, Joel W. Ager*

1. **Synthesis of spray pyrolyzed thin films**

   Iron pyrite thin films were fabricated simple spray pyrolysis method. In a typical synthesis procedure, Iron chloride $FeCl_3.6H_2O$ (Sigma Aldrich) (Fe precursor) and thiourea $NH_2CSNH_2$ (Sigma Aldrich) (sulfur precursor) were mixed in a certain ratio in aqueous solution and sprayed on a hot surface to react and form film. 0.1 M solution of iron chloride (1.35 g in 50 ml of DI water) and thiourea were first prepared (0.38 g in 50 ml of DI water). The two solutions were mixed in more DI water in Fe:S atomic ratio of 1:6 by weight to get the spraying solution. For example, a total of 50 ml spraying solution contains 7 ml of 0.1 M $FeCl_3.H_2O$ solution, 4.2 ml of 0.1 M thiourea solution and 38.8 ml of DI water. Thin films were deposited on a quartz substrate placed on a hot plate at 350 °C. As-deposited films were further heat treated in sulfur ambient to get crystalline and phase pure pyrite films. Sulfurization was performed at atmospheric pressure in an open gas flow quartz tube system under moderate Argon gas flow rate of 15 sccm. Before sulfurization, system was pumped down to remove oxygen from the system and later purged with Argon 3-4 times. During sulfurization 100 mg of sulfur powder was kept on a ceramic boat at around 300 °C and substrate was kept at 400 °C at the centre of the zone and the process carried out for 30 minutes after which all the sulfur powder was completely consumed. Source (sulfur boat) and substrate were 12 cm apart. Quartz tube used for sulfurization had inner and outer diameter of 21 mm and 25 mm respectively.

2. **Synthesis of nanocubes by hot-injection method**

Anhydrous 98% Iron (II) chloride ($FeCl_2$), 70 % oleylamine (OLA), and sulfur powder from Sigma-Aldrich. 0.5 mmol (63.5 mg) of $FeCl_2$ was mixed with 5 mL OLA in a tri-neck flask and degassed for 30 minutes. Subsequently it was heated to 110 °C for 1 hour to form the Fe-OLA complex. Thereafter, the temperature was raised to 180 °C and 3 mmol (96 mg) of sulfur dissolved in OLA was injected into the flask. The temperature was maintained at 180 °C for 24 hours to complete the reaction. After reaction, the solution was allowed to cool to room temperature naturally and a large amount of methanol was added to precipitate the $FeS_2$ nanocubes. The product was concentrated by centrifugation and then dispersed in hexane. 50 µL of Pyrite NCs suspension with a concentration of (0.1M) was spin-coated on a quartz substrate sequentially to obtain a pyrite film and subsequently sulfurized under in a quartz tube at 500 °C for 30 minutes. The obtained films were carried out for electrical measurements.

3. **Pulsed laser deposition**

The chamber was evacuated to a base pressure below $1.8 \times 10^{-7}$ torr. The cleaned and dried substrate was mounted on the stage and was introduced into the growth chamber through a load lock chamber and its temperature was increased to 100 °C with a constant ramp rate of 10 °C/min by a stage heater. The chamber pressure was then relieved to the growth pressure which was typically in the range $10^{-6}$ -$10^{-3}$ torr via a gate valve. Deposition at the low pressure range of ~ $4.2 \times 10^{-6}$ torr was done for 12 minutes without using any background gas. The laser source, a Coherent COMPexPro pulsed KrF excimer laser operating at 284 nm, was set at 10 ns pulse width, 10 Hz repetition rate, and 150 mJ pulse energy. The beam was directed through a spherical lens and a slit, focused to a 1.5 cm line, and swept across the target in raster scan manner. Substrate to target distance was kept about 12 cm. The substrate was continuously rotated at 2 rpm to ensure uniformity of deposition. In situ film thickness was monitored with an SQM-160 quartz crystal microbalance. Films were subsequently separately sulfurized under thermal and H$_2$S plasma sulfurization.

In the thermal sulfurization process, Sulfur powder was kept at an upstream point in the tube while the film was at the center of the tube at 400 °C for 2 hrs under constant N$_2$ gas flow of 20 sccm). The quantity of sulfur powder used was around 500 mg for each heat treatment to ensure the adequacy of sulfur to last throughout the duration of the process. For plasma sulfurization, an inductively coupled plasma source was used to generate the sulfur plasma from H$_2$S gas. The films were exposed to the H$_2$S plasma for 60 minutes to obtain iron sulfide thin films.

## 4. Density Functional Theory

Density functional theory (DFT) calculations were performed using the plane-wave code Vienna Ab-initio Simulation Package (VASP).[1] Electron-ion interactions were described by the Projector augmented wave (PAW) method.[2,3] Exchange and correlation was described by generalized gradient approximation as parameterized by Perdew-Burke-Ernzerhof (PBE). The kinetic energy cutoff for plane waves was set to 350 eV. The atomic positions were considered to be optimized when the forces on each ion were less than 0.01 eV/Å. A Hubbard U correction of 1.8 eV was applied to Fe 3d orbitals.[4,5] Iron Pyrite is a low-spin semiconductor. This was also confirmed by spin polarized DFT calculations (The low-spin configuration was verified to be the ground state within PBE and PBE+U. For bulk reference calculations, Monkhorst-Pack set of k-points[6] of 8×8×8 and 4×4×4 were used for primitive unit cell and 2x2x2 supercell, respectively for geometry optimizations. For density of states (DOS) calculations, a dense k-point mesh of 11×11×11 was employed.

## 5. Electrical transport, structural and optical measurements

Electrical transport measurements were performed using PPMS (Quantum Design). Seebeck measurements were performed using a home-built set-up described elsewhere.[7] SEM images were acquired using JEOL JSM-7600F field-emission scanning electron microscope. T64000 micro-Raman spectrometer with an incident power of 0.14 mW and laser illumination of 532 nm wavelength was used for Raman analysis.

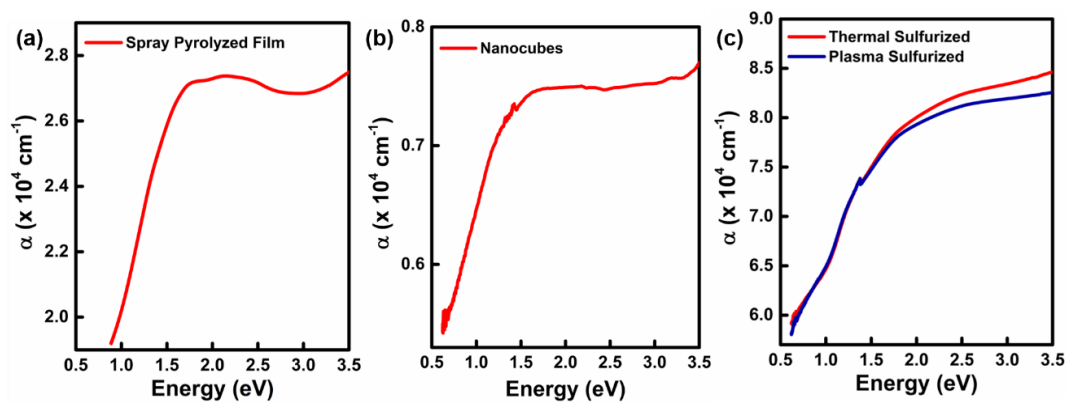

**Figure S1**. Optical absorption spectra of (a) spray pyrolyzed and sulfurized, (b) spin coated nanocubes and sulfurized and (c) PLD deposited thermally and $H_2S$ plasma sulfurized thin film.

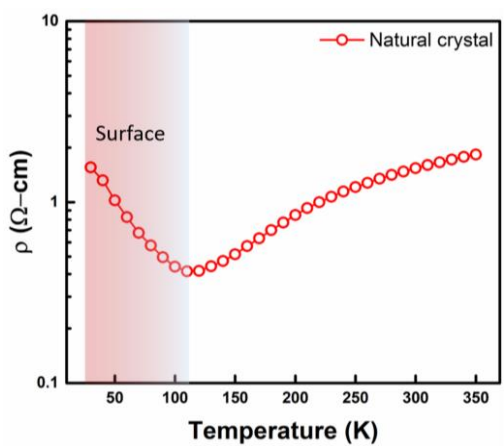

**Figure S2.** Electronic charge transport measurement performed on pyrite single crystal. Temperature dependence of resistivity and region of dominant surface conduction marked in shaded colour.

6. **Linearization of the transport equation**

From eq. (1), $\rho = \rho_o * exp\,(T_o/T)^\alpha$

$$\ln(\rho) = \ln(\rho_o) + \left(\frac{T_o}{T}\right)^\alpha$$

Substitute, $T = e^x$

To Calculate, $d(\ln \rho)/d(\ln T)$

$$\ln(\rho) = \ln(\rho_o) + T_o^\alpha * e^{-\alpha x}$$

$$d(\ln \rho)/dx = -\alpha\, T_o^\alpha e^{-\alpha x}$$

$$d(\ln \rho)/d(\ln T) = -\alpha \left(\frac{T_o}{T}\right)^\alpha$$

$$\ln\left(-\frac{d\,(\ln \rho)}{d\,(\ln T)}\right) = \ln(\alpha) + \ln\left(\frac{T_o}{T}\right)^\alpha$$

$$\ln\left(-\frac{d\,(\ln \rho)}{d\,(\ln T)}\right) = \ln(\alpha) + \alpha \ln(T_o) + \alpha \ln(T)$$

$$\ln\left(-\frac{d\,(\ln \rho)}{d\,(\ln T)}\right) = Constant - \alpha \ln(T)$$

$$\ln(W) = Constant - \alpha \ln(T)$$

Where, $$W = -\frac{d\,(\ln \rho)}{d\,(\ln T)}$$

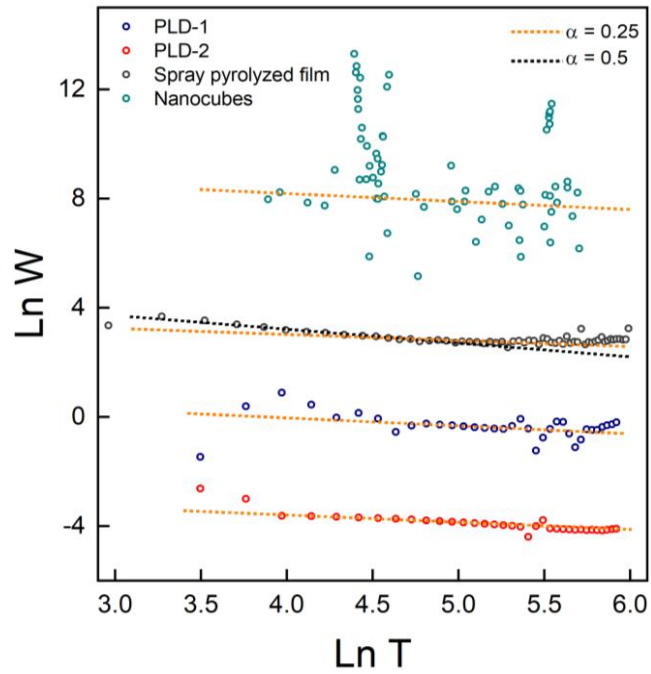

**Figure S3.** Linearized curve (Zabrodskii plot) for the spray pyrolyzed film, slopes corresponding to exponent α = 0.25 and 0.5 are shown for reference. The nanocubes had higher resistivity than the other films, which increases the uncertainty for this type of analysis. In this case, the dotted line is shown to serve as a reference for the slope corresponding to the exponent (the exponent is not determined from the fitting).

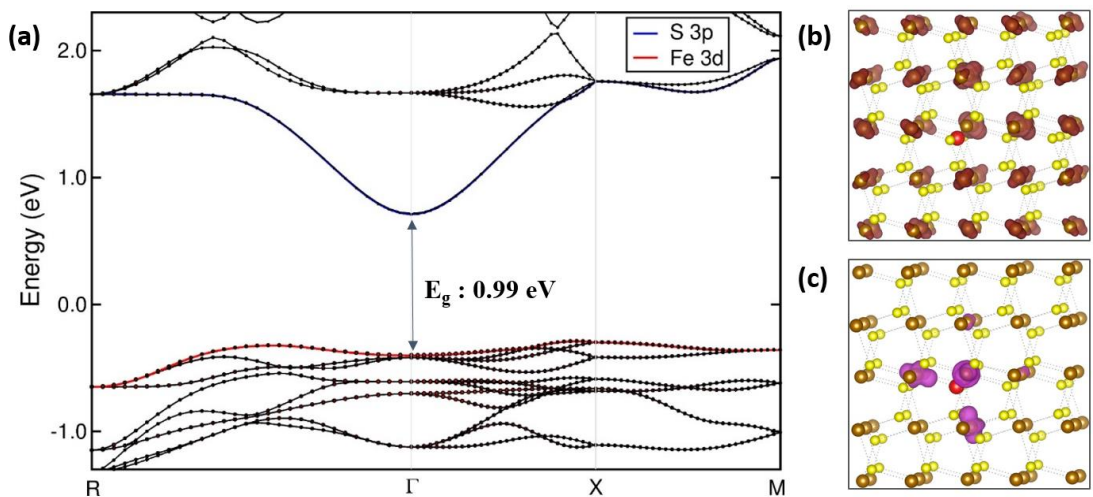

**Figure S4.** (a) Calculated density of states (DOS) and the bandgap. Isosurface ($\rho$=0.015e/Bohr$^3$) of decomposed charge density in a 2×2×2 unit cell containing a SV

for (b) VBM and (c) CBM. The location of SV is shown as a red colored sphere. Brown and yellow colored spheres represent Fe and S atoms, respectively.

From the band decomposed charge density figures, we can see that the defect states are mainly contributed by atoms close to the defect only. The states near to VBM are mainly contributed by three Fe atoms close to the defect but which were not bonded to the removed S atoms while the defect state close to CBM is contributed by three Fe atoms which have broken bonds, i.e. these Fe atoms had bonds with the removed S atom. The reason for electron density visible mainly on Fe atoms is mainly because there are more number of valence electrons in Fe atom's pseudopotential than in S atom's.

Table S1. Carrier density and mobility values for various $FeS_2$ single crystals, thin films and nanostructures from the literature including the current work.

| $FeS_2$ | Carrier density ($cm^{-3}$) | Carrier mobility ($cm^2/V\text{-}s$) | Carrier type | Ref. |
|---|---|---|---|---|
| **Single crystal** | $2.0 \times 10^{16}$ | 280 | n-type | [8] |
| **Single crystal** | $2.1 \times 10^{16}$ | 366 | n-type | [8] |
| **Single crystal** | $1.5 \times 10^{16}$ | 230 | n-type | [8] |
| **Single crystal** | $5.0 \times 10^{15}$ | 245 | n-type | [9] |
| **Single crystal** | $1.0 \times 10^{16}$ | 185 | n-type | [10] |
| **Single crystal** | $1.0 \times 10^{15}$ | 50 | n-type | [11] |
| **Single crystal** | $1.2 \times 10^{17}$ | 164 | n-type | [12] |
| **Single crystal** | $2.0 \times 10^{15}$ | 40 | n-type | [13] |
| **Single crystal** | $2.0 \times 10^{17}$ | 200 | n-type | [13] |
| **Single crystal** | $5.1 \times 10^{16}$ | 64 | n-type | This work |
| **Thin film** | $5.0 \times 10^{18}$ | 5 | p-type | [14] |
| **Thin film** | $3.0 \times 10^{17}$ | 14 | p-type | [15] |
| **Thin film** | $4.8 \times 10^{18}$ | 2.12 | p-type | [16] |
| **Thin film** | $1.0 \times 10^{19}$ | - | p-type | [17] |
| **Thin film** | $5.5 \times 10^{17}$ | 280 | p-type | [18] |
| **Thin film** | $1.0 \times 10^{20}$ | 25 | p-type | [19] |
| **Thin film** | $1.0 \times 10^{21}$ | 1 | n-type/p-type | [20] |
| **Thin film** | $1.0 \times 10^{19}$ | 10 | n-type | [21] |
| **Thin film** | $7.0 \times 10^{18}$ | 1.4 | p-type | This work |
| **Thin film** | $4.0 \times 10^{19}$ | 0.3 | p-type | This work |
| **Thin film** | $2.7 \times 10^{18}$ | 4.2 | p-type | This work |
| **Nanostructure** | $2.4 \times 10^{21}$ | 0.08 | p-type | [22] |
| **Nanostructure** | $1.0 \times 10^{21}$ | 0.03 | p-type | [23] |
| **Nanostructure** | $1.0 \times 10^{19}$ | 0.2 | p-type | [24] |
| **Nanostructure** | $4.9 \times 10^{18}$ | 0.3 | p-type | [25] |
| **Nanostructure** | $5.3 \times 10^{18}$ | 0.1 | p-type | This work |

**Supplemental references**

(1) Kresse, G.; Furthmüller, J. Efficient iterative schemes for ab initio total-energy calculations using a plane-wave basis set. Phys. Rev. B 1996, 54, 11169–11186.

(2) Blöchl, P. E. Projector augmented-wave method. Phys. Rev. B 1994, 50, 17953–17979.

(3) Kresse, G.; Joubert, D. From ultrasoft pseudopotentials to the projector augmented-wave method. Phys. Rev. B 1999, 59, 1758–1775.

(4) Dudarev, S. L.; Botton, G. A.; Savrasov, S. Y.; Humphreys, C. J.; Sutton, A. P. Electron energy- loss spectra and the structural stability of nickel oxide: An LSDA+U study. Phys. Rev. B 1998, 57, 1505–1509.

(5) Perdew, J. P.; Wang, Y. Accurate and simple analytic representation of the electron-gas correlation energy. Phys. Rev. B 1992, 45, 13244–13249.

(6) Monkhorst, H. J.; Pack, J. D. Special points for Brillouin-zone integrations. Phys. Rev. B 1976, 13, 5188–5192.

(7) Ager, J. W. *et al.* Mg-doped InN and InGaN – photoluminescence, capacitance–voltage and thermopower measurements. Physica Status Solidi, 2008, 245, 873–877.

(8) Schieck, R., *et al.*, Electrical properties of natural and synthetic pyrite (FeS2) crystals. Journal of Materials Research, 1990. 5(7): p. 1567-1572.

(9) Limpinsel, M., *et al.*, An inversion layer at the surface of n-type iron pyrite. Energy & Environmental Science, 2014. 7(6): p. 1974-1989.

(10) Willeke, G., *et al.*, Preparation and electrical transport properties of pyrite (FeS2) single crystals. Journal of Alloys and Compounds, 1992. 178(1): p. 181-191.

(11) Cabán-Acevedo, M., *et al.*, Ionization of High-Density Deep Donor Defect States Explains the Low Photovoltage of Iron Pyrite Single Crystals. Journal of the American Chemical Society, 2014. 136(49): p. 17163-17179.

(12) Echarri, A.L. and C. Sánchez, "n" type semiconductivity in natural single crystals of FeS2 (pyrite). Solid State Communications, 1974. 15(5): p. 827-831.

(13) Voigt, B., *et al.*, Transport Evidence for Sulfur Vacancies as the Origin of Unintentional n-Type Doping in Pyrite FeS2. ACS Applied Materials & Interfaces, 2019. **11**(17): p. 15552-15563.

(14) Willeke, G., *et al.*, Thin pyrite (FeS2) films prepared by magnetron sputtering. Thin Solid Films, 1992. 213(2): p. 271-276.

(15) Mazón-Montijo, D.A., M.T.S. Nair, and P.K. Nair, Iron Pyrite Thin Films *via* Thermal Treatment of Chemically Deposited Precursor Films. ECS Journal of Solid State Science and Technology, 2013. 2(11): p. P465-P470.

(16) Shukla, S., *et al.*, Iron Pyrite Thin Film Counter Electrodes for Dye-Sensitized Solar Cells: High Efficiency for Iodine and Cobalt Redox Electrolyte Cells. ACS Nano, 2014. 8(10): p. 10597-10605.

(17) Berry, N., *et al.*, Atmospheric-Pressure Chemical Vapor Deposition of Iron Pyrite Thin Films. Advanced Energy Materials, 2012. 2(9): p. 1124-1135.

(18) Takahashi, N., *et al.*, Growth of Single-Crystal Pyrite Films by Atmospheric Pressure Chemical Vapor Deposition. Chemistry of Materials, 2003. 15(9): p. 1763-1765.

(19) Lichtenberger, D., *et al.*, Structural, optical and electrical properties of polycrystalline iron pyrite layers deposited by reactive d.c. magnetron sputtering. Thin Solid Films, 1994. 246(1): p. 6-12.

(20) Zhang, X., *et al.*, Phase Stability and Stoichiometry in Thin Film Iron Pyrite: Impact on Electronic Transport Properties. ACS Applied Materials & Interfaces, 2015. 7(25): p. 14130-14139.

(21) Zhang, X., *et al.*, Potential resolution to the doping puzzle in iron pyrite: Carrier type determination by Hall effect and thermopower. Physical Review Materials, 2017. 1(1): p. 015402.

(22) Liang, D., *et al.*, Gated Hall Effect of Nanoplate Devices Reveals Surface-State-Induced Surface Inversion in Iron Pyrite Semiconductor. Nano Letters, 2014. 14(12): p. 6754-6760.

(23) Cabán-Acevedo, M., *et al.*, Synthesis and Properties of Semiconducting Iron Pyrite (FeS2) Nanowires. Nano Letters, 2012. 12(4): p. 1977-1982.

(24) Cabán-Acevedo, M., *et al.*, Synthesis, Characterization, and Variable Range Hopping Transport of Pyrite (FeS2) Nanorods, Nanobelts, and Nanoplates. ACS Nano, 2013. 7(2): p. 1731-1739.

(25) Kinner, T., *et al.*, Majority Carrier Type Control of Cobalt Iron Sulfide (CoxFe1–xS2) Pyrite Nanocrystals. The Journal of Physical Chemistry C, 2016. 120(10): p. 5706-5713.